\documentclass[12pt, JPCM]{iopart}
\usepackage{iopams}  
\usepackage{hyperref}
\usepackage{graphicx}
\usepackage[numbers,sort&compress]{natbib}

\hypersetup{hypertex=true,
	colorlinks=true,
	linkcolor=blue,
	urlcolor = blue,
	citecolor=blue}

\begin{document}

\title{Nonlocal pseudospin dynamics in a quantum Ising chain}
\author{K L Zhang and Z Song*}
\address{School of Physics, Nankai University, Tianjin 300071, China}
\begin{indented}
	\item[]*Author to whom any correspondence should be addressed.
\end{indented}
\ead{songtc@nankai.edu.cn}
\vspace{10pt}
\begin{indented}
	\item[]\today
\end{indented}

\begin{abstract}
The existence of topological zero modes in nontrivial phase of quantum Ising
chain results in not only the Kramers-like degeneracy spectrum, but also
dynamic response for non-Hermitian perturbation in the ordered phase  
(2021 \textit{Phys. Rev. Lett.} \textbf{126} 116401).  
In this work, we investigate the possible response of the degeneracy spectrum for Hermitian
perturbations. We provide a single-particle description of the model in the
ordered phase, associating with an internal degree of freedom characterized
as a pseudospin. The effective magnetic field, arising from both local and
nonlocal perturbations in terms of string operators, acts on the
pseudospin. We show that the action of string operator can be realized via a
quench under the local perturbations. As an application, any ground states
and excited states for the Hamiltonian with perturbation can be selected to
identify the quantum phase, by adding the other perturbations to trigger a
quench and measuring the Loschmidt echo.
\end{abstract}
\noindent{\it Keywords\/}: transverse field Ising chain, quench dynamics, nonlocal pseudospin, quantum phase transition, thermal state
\maketitle

\section{Introduction}

Identifying the quantum phase diagram of a physical system is of vital
importance in both condensed matter physics and quantum information science.
In the past few decades, a large number of theoretical and experimental
studies in this field have emerged~\cite{fisher1990quantum, bitko1996quantum, vojta2000quantum, si2001locally, porras2004effective, uhlarz2004quantum, ronnow2005quantum, coldea2010quantum, kim2011quantum, simon2011quantum, trenkwalder2016quantum, rem2019identifying}. The
transverse field Ising chain~\cite{pfeuty1970one,sachdev2011quantum,dutta2015quantum}, which is one of the paradigmatic model to explore quantum
phase transition (QPT) and quantum information science, plays a key role in
this realm. Experimentally, the transverse field Ising chain has direct 
	realizations \cite{bitko1996quantum, coldea2010quantum}, 
	and is also achievable with ultracold atoms in optical 
	lattices \cite{simon2011quantum, islam2011onset}. In this model, the integrability has possibly attracted the 
attention of the researchers the most. In spite of the simplicity, it
possesses all the basic elements of the QPT. The competition between the
nearest neighbor interaction and the external magnetic field leads to an
ordered phase and a disordered phase, which are separated by a quantum
critical point, accompanied by a spontaneous symmetry breaking. 

Methodologically, triggering the quantum quench dynamics \cite%
{polkovnikov2011colloquium, essler2016quench, abeling2016quantum,
jurcevic2017direct, granet2020finite} by suddenly changing system parameters
is frequently used to study the QPT. After the quench, the system undergoes
nonequilibrium dynamics \cite{polkovnikov2011colloquium}, and one can
measure the Loschmidt echo (LE) \cite{andraschko2014dynamical,
quan2006decay, cozzini2007quantum, heyl2013dynamical, jafari2017loschmidt,
mera2018dynamical} to quantify the deviation of the evolved state from the
initial state. Generally speaking, since the evolved state contains 
information of both the initial state and the postquench Hamiltonian, the
behavior of LE can reflect the physical properties of the system. Besides
the application in QPT, it is interesting to employ the quench protocol to
manipulating the spin degrees of freedom in quantum information science.
Experimentally, the observation of quench dynamics in a quantum Ising chain have been realized in a trapped-ion quantum simulator \cite{jurcevic2017direct}, where the ion chain is initialized in a ground state of the prequench Hamiltonian with zero magnetic field and suddenly apply a large magnetic field.

In this paper, we investigate the possible response of the degeneracy
spectrum of the transverse field Ising chain to Hermitian perturbations, 
which is more convenient for experimental implementation than a non-Hermitian one \cite{zhang2021quantum}. A non-Hermitian Hamiltonian, which effectively describes the dynamics at short time for an open quantum system \cite{lee2014heralded, bergholtz2021exceptional, roccati2022non}, is more complicated to be considered in experiment, since the form of non-Hermitian term depends on the coupling to the environment. Also, at a longer time, the effect of decoherence must be considered.  
On the other hand, most of the studies about quench dynamics focused on the ground state and some simple state, such as saturated ferromagnetic state, as the initial states. We tried to extend the initial states to any excited states, by utilized the identical split of all energy levels of the system under certain perturbation.
We focus on the model with open boundary condition. 
It is shown that \cite{zhang2021quantum} the Kramers-like degeneracy spectrum in the ordered phase is related to the topological zero modes of the Kitaev chain \cite{kitaev2001unpaired}, thus the degeneracy is topologically protected, in other words, robust against certain local perturbations. However, it is not easy to manipulate such an inner degree of 
freedom, as well as breaking the degeneracy with arbitrary local perturbation. Our
motivations are to provide a quench protocol to identify the topology-related  degeneracy spectrum and the quantum phases, and seek possible action to manipulate the inner degree of freedom for this model with Hermitian
perturbations that result in the identical split in the spectrum. To
this end, we provide a single-particle description of the model in the
ordered phase, associating with an internal degree of freedom characterized
as a pseudospin. The effective magnetic field acts on the pseudospin,
which arises from both local and nonlocal perturbations in terms of string
operators.  As an application, a quantum state living in the two-fold
degenerate subspace of the system in the ordered phase can be considered as
a single qubit, and with the effective magnetic field, the time evolution
operator can achieve two quantum gates---phase gate and Hadamard gate. To our knowledge, there is no direct experimental realization of the string operators, although some multi-site interacting terms have been discussed in other theoretical spin models \cite{suzuki1971relationship, divakaran2013three}. 
Thus, we provide a scheme to realize the actions of the string operators discussed in this paper. We
simplify the perturbations and show that the string operator action can be realized via a quench
under the local perturbations. Any ground state or excited state for the
Hamiltonian with perturbation can be selected for identifying the quantum
phase, by adding the other perturbations to trigger a quench and measuring
the LE. A possible application in the thermal state is also discussed. It is worth noting that most of the studies about QPT and quench dynamics focused on the ground state of the system. In
addition, numerical simulations for a finite-size system are provided to
support our results.

The remainder of this paper is organized as follows. In section \ref{II}, we
present the transverse field Ising chain and its symmetries. In section \ref%
{III}, we introduce the pseudospin description for the model in the ordered
phase and investigate the response of the degenerate spectrum to
perturbations. In section \ref{IV}, we discuss the simplification of the perturbations, the realization of the nonlocal operator, and the possible applications in the QPT and thermal state with
numerical results of LEs. Finally, we summarize our results in section \ref{V}.

\section{Model and symmetries}

\label{II}

In this section, we present the Hamiltonian and a brief review on its basic
properties, based on which we perform our investigations in this work. The
model considered is the transverse field Ising chain with open boundary
condition, defined by the Hamiltonian 
\begin{equation}
H_{0}=-J\sum_{j=1}^{N-1}\sigma _{j}^{x}\sigma
_{j+1}^{x}+g\sum_{j=1}^{N}\sigma _{j}^{z},  \label{H_Ising}
\end{equation}%
where $\sigma _{j}^{\alpha }$ ($\alpha =x,$ $y,$ $z$) are the Pauli
operators on site $j$ and parameter $g$ ($g > 0$) is the transverse field
strength. For simplicity, the following discussion assumes that $J=1$. It
can be checked that the model respects two symmetries. The first one is the
parity symmetry, that is, the parity operator $p=\prod_{j=1}^{N}(-\sigma _{j}^{z})$ is commutative with 
the Hamiltonian. The second one is a little subtle and is crucial to our main conclusion 
\cite{zhang2021quantum, zhang2020ising}. The model with periodic boundary
condition is exactly solvable and has been well studied \cite{pfeuty1970one}%
. At zero temperature, the QPT at $g=1$ separates an ordered phase of the
system ($g<1$) from a disordered phase ($g>1$). However, when we consider
the model with open boundary condition, it possesses an exclusive symmetry
in the ordered phase $g<1$ in thermodynamic limit. Defining a nonlocal
operator 
\begin{equation}
D=\frac{1}{2}\sqrt{1-g^{2}}\sum_{j=1}^{N}g^{j-1}\left[\prod\limits_{l<j}\left( -\sigma _{l}^{z}\right) \sigma
_{j}^{x}-\rmi\prod\limits_{l<N-j+1}\left( -\sigma _{l}^{z}\right) \sigma
_{N-j+1}^{y}\right]  ,  \label{D}
\end{equation}%
(where $\rmi=\sqrt{-1}$), we have $D$ and $D^{\dagger}$ commute with the Hamiltonian \cite{zhang2021quantum} , 
which is referred to as edge-spin symmetry, since {it is the outcome of the
edge operator of the Kitaev chain \cite{kitaev2001unpaired}. Operator }$D$
is a fermion operator, obeying the relations $\{D,D^{\dag }\}=1$ and $%
D^{2}=(D^{\dag })^{2}=0$. It should be noted that it is contingent on the
following conditions: $g<1$, a large $N$ limit, and open boundary. In
addition, operator $D$ is non-universal and $g$-dependent.

From these symmetries, we have following implications: the complete
eigenstates $\left\{ \left\vert \psi _{n}^{+}\right\rangle ,\left\vert \psi
_{n}^{-}\right\rangle \right\} $ of $H_{0}$ with eigenenergy $\varepsilon
_{n}^{\pm }$, $H_{0}\left\vert \psi _{n}^{\pm }\right\rangle =\varepsilon
_{n}^{\pm }\left\vert \psi _{n}^{\pm }\right\rangle $, span two invariant
subspaces for any value of $g$, where $\pm$ denotes the eigenvalues of parity operator $p$. Importantly, within the region $g<1 $, 
 the edge-spin symmetry guarantees the existence of eigenstates degeneracy $\varepsilon
_{n}^{+}=\varepsilon _{n}^{-}=\varepsilon _{n}$, referred to as Kramers-like
degeneracy. Accordingly, we also have the relations 
\begin{equation}
\rmi\left( D^{\dag }-D\right) \left( \left\vert \psi _{n}^{+}\right\rangle \pm
\left\vert \psi _{n}^{-}\right\rangle \right) =\pm \rmi\left( \left\vert \psi
_{n}^{+}\right\rangle \mp \left\vert \psi _{n}^{-}\right\rangle \right) ,
\end{equation}%
and%
\begin{equation}
p\left( \left\vert \psi _{n}^{+}\right\rangle \pm \left\vert \psi
_{n}^{-}\right\rangle \right) =\left\vert \psi _{n}^{+}\right\rangle \mp
\left\vert \psi _{n}^{-}\right\rangle ,
\end{equation}%
which play an important role in the quench dynamics, as demonstrated in the
following section.

\section{Nonlocal pseudospin and Loschmidt echo}

\label{III}

It is not surprising that a Hermitian perturbation can lift the degeneracy.
However, it should lead to a fascinating dynamic phenomenon if an identical
split in each level in the spectrum is obtained, which enable the same oscillatory dynamics of the excited states as that of the ground state. Moreover, hybridizing two
robust degenerate states on demand is a central task of quantum information
processing, since these states are immune to weak local perturbations.

We first focus on the ordered quantum phase $0<g<1$, considering a perturbed
Hamiltonian%
\begin{equation}
H=H_{0}+H^{\prime },
\label{H_perturbed}
\end{equation}%
with $H^{\prime }$ being the combination of three types of actions, 
\begin{equation}
H^{\prime }=\kappa _{x}(D^{\dag }+D)+\rmi\kappa _{y}(D^{\dag }-D)+\kappa _{z}p.
\label{Hp}
\end{equation}%
We note that the perturbation is nonlocal, containing the string operators $%
\prod\nolimits_{l}\left( -\sigma _{l}^{z}\right) $. The perturbed Hamiltonian in equation  (\ref{H_perturbed}) can be changed to the fermionic form by Jordan-Wigner transformation \cite{jordan1993paulische}. In this representation $H^{\prime }$ only contains the local terms, however, this is impractical in experimental aspect since $H^{\prime }$ break the fermionic parity of the system \cite{kitaev2001unpaired, wick1997intrinsic, bradler2012comment}. In contrast, the parity symmetry of a spin system does not need to be conserved.

Any pair of degenerate eigenstates $\left( \left\vert \psi _{n}^{+}\right\rangle ,\left\vert \psi_{n}^{-}\right\rangle \right) $ with energy $\varepsilon _{n}$ spans a
diagonal block in the form%
\begin{equation}
\mathbf{B} \cdot \boldsymbol{\sigma }=\left( 
\begin{array}{cc}
\kappa _{z} & \kappa _{x}-\rmi\kappa _{y} \\ 
\kappa _{x}+\rmi\kappa _{y} & -\kappa _{z}%
\end{array}%
\right) ,
\end{equation}%
with parameter vector $\mathbf{B}=(\kappa _{x},\kappa _{y},\kappa _{z})$ and
Pauli matrix $\boldsymbol{\sigma }=(\sigma _{x},\sigma _{y},\sigma _{z})$.
Therefore, under the basis $\left( \left\vert \psi _{1}^{+}\right\rangle
,\left\vert \psi _{1}^{-}\right\rangle ,\left\vert \psi
_{2}^{+}\right\rangle ,\left\vert \psi _{2}^{-}\right\rangle ,...\right) $,
we get an equivalent Hamiltonian for $H$ 
\begin{equation}
H_{\mathrm{eq}}=\bigoplus_{n=1}^{2^{N-1}}\left( \mathbf{B} \cdot \boldsymbol{%
\sigma }+\varepsilon _{n}I_{2}\right) ,
\end{equation}%
where $I_{2}$ denotes the $2\times2$ identity matrix. In this single-particle description with pseudo spin, it is obvious that the present of $\mathbf{B}$ splits the degeneracy of the energy levels $\varepsilon_{n}$. This is not contradictory with the previous claim about the degeneracy of the energy spectrum is robust.  
The degeneracy is robust against the random variations on the uniform distribution of parameters $(J,\ g)$ in $H_{0}$. This can be proved by the robustness of the edge-spin symmetry: it can be checked that the commutation relation $[D,\ H_{0}]=0$ still hold for the system with local perturbation on system parameters in large $N$ limit \cite{zhang2020dynamic}, leading to the degeneracy that is robust against local perturbation. This means that when site-dependent $(J_{j},\ g_{j})$ are disordered and the field $\mathbf{B}$ is zero, a qubit is stable in the degenerate subspace, without the influence of dynamics phase factor. 

The introduction of the perturbation term $H^{\prime }$ allows two possible
applications through time evolution. First, the operation on the robust
degenerate state can be realized. We stress that this is not an outcome of any Hermitian perturbations lifting degeneracy. It needs three independent Hermitian perturbations to realize the full operation on the robust degenerate state. In the degenerate subspace with index $n$,
an arbitrary state $\left\vert \psi _{n}\right\rangle =\alpha \left\vert
\psi _{n}^{+}\right\rangle +\beta \left\vert \psi _{n}^{-}\right\rangle $
acts as a single qubit, where $\alpha $ and $\beta $ are complex numbers
encoding quantum information. An arbitrary unitary operation on this state
can be realized by the time evolution operator 
\begin{eqnarray}
U\left( t\right)  &=&\rme^{-\rmi\varepsilon _{n}t}\rme^{-\rmi\mathbf{B}\cdot \boldsymbol{%
\sigma }t}  \nonumber \\
&=&\rme^{-\rmi\varepsilon _{n}t}\left[ \cos \left( \left\vert \mathbf{B}%
\right\vert t\right) -\rmi\frac{\mathbf{B}\cdot \boldsymbol{\sigma }}{%
\left\vert \mathbf{B}\right\vert }\sin \left( \left\vert \mathbf{B}%
\right\vert t\right) \right] ,
\end{eqnarray}%
by choosing appropriate parameters $\mathbf{B}=(\kappa _{x},\kappa
_{y},\kappa _{z})$ and evolved time $t$. For example, when choosing $\mathbf{%
B}=(0,0,\kappa _{z})$, it realizes the action of the phase gate 
\begin{equation}
\mathcal{P}=\rme^{\rmi\varepsilon _{n}t+\rmi\kappa _{z}t}U\left( t\right) =\left( 
\begin{array}{cc}
1 & 0 \\ 
0 & \rme^{2\rmi\kappa _{z}t}%
\end{array}%
\right) .
\end{equation}%
When $\mathbf{B}=(\kappa _{x},0,\kappa _{x})$ and $t=t^{\prime }=\pi
/(2\left\vert \mathbf{B}\right\vert )$, we get the Hadamard gate 
\begin{equation}
\mathcal{H}=\rmi\rme^{\rmi\varepsilon _{n}t^{\prime }}U\left( t^{\prime }\right) =%
\frac{1}{\sqrt{2}}\left( 
\begin{array}{cc}
1 & 1 \\ 
1 & -1%
\end{array}%
\right) .
\end{equation}

Second, when considering the LE of quench dynamics in the ordered phase, the oscillation behavior of trigonometric function can be observed. In the next section, a $g$-independent form of perturbation is derived from $H'$, which is also valid in the disordered phase and can be utilized for identifying the quantum phases. In the following, we only give an analytical analysis of the expected results
with parameter in the ordered quantum phase, since $H^{\prime }$ is $g$%
-dependent and is only defined in the region $0<g<1$. It can be checked that the solution of $%
H_{\mathrm{eq}}$\ is simply given by the eigenvectors of $\mathbf{B} \cdot 
\boldsymbol{\sigma } $, that is,%
\begin{eqnarray}
\left\vert \phi _{n}^{+}\right\rangle &=&\cos \frac{\theta }{2}\left\vert
\psi _{n}^{+}\right\rangle +\sin \frac{\theta }{2}\rme^{\rmi\varphi }\left\vert
\psi _{n}^{-}\right\rangle , \label{phip}\\
\left\vert \phi _{n}^{-}\right\rangle &=&\sin \frac{\theta }{2}\left\vert
\psi _{n}^{+}\right\rangle -\cos \frac{\theta }{2}\rme^{\rmi\varphi }\left\vert
\psi _{n}^{-}\right\rangle ,\label{phim}
\end{eqnarray}%
with corresponding eigenenergies 
\begin{equation}
E_{n}^{\pm }=\varepsilon _{n}\pm \left\vert \mathbf{B}\right\vert
=\varepsilon _{n}\pm \sqrt{\kappa _{x}^{2}+\kappa _{y}^{2}+\kappa _{z}^{2}},
\end{equation}%
where $\theta $ and $\varphi $ are defined as $\cos \theta =\kappa
_{z}/\left\vert \mathbf{B}\right\vert $ and $\tan \varphi =\kappa
_{y}/\kappa _{x}$. 
The splitting of energy levels can be equivalently understood as a set of two-level atoms in a Zeeman magnetic field, which breaks parity symmetry of the system when $\kappa_{x}\neq0$ or $\kappa_{y}\neq0$, which is clear from the eigenvectors in equations (\ref{phip}) and (\ref{phim}). This result is significant for constructing the prequench and postquench Hamiltonians, as well as for understanding the oscillatory dynamics.
The level spacing $2\sqrt{\kappa _{x}^{2}+\kappa
_{y}^{2}+\kappa _{z}^{2}}$\ is $n$ and $g$ independent, resulting in a
periodic dynamics among all the spectrum. It is exclusive and therefore, is the
manifestation of the ordered quantum phase.

To capture the effect of perturbation $H^{\prime }$ on the dynamics, we
introduce the LE, which is a measure of reversibility and sensitivity to
perturbation of quantum evolution. An initial quantum state $\left\vert \Phi
(0)\right\rangle $ evolves during a time $t$ under a postquench Hamiltonian $%
H_{\rm Pos}$ reaching state $\left\vert \Phi (t)\right\rangle $.
The behavior of $\langle \Phi (0)\left\vert \Phi (t)\right\rangle $ is the
basis of LE measurement, in which $\left\vert \Phi (0)\right\rangle $\ is an
eigenstate of prequench Hamiltonian $H_{\rm Pre}$. Two
Hamiltonians $H_{\rm Pre}$\ and $H_{\rm Pos}$\ can be
taken by two different sets of parameters $\left( \kappa _{x},\kappa
_{y},\kappa _{z}\right) $ for $H$. The LE is defined as%
\begin{equation}
L\left( t\right) =|\left\langle \Phi (0)\right\vert \rme^{\rmi H_{\rm Pos}%
t}\rme^{-\rmi H_{\rm Pre}t}\left\vert \Phi (0)\right\rangle |^{2},
\end{equation}%
where $\left\vert \Phi (0)\right\rangle $ is usually an easily prepared
state, such as the ground state of $H_{\rm Pre}$.\ The dynamics
in each subspace is clearly a rotation of a Bloch state with a fixed axis
and frequency. Then a maximum oscillating amplitude is achieved when the
initial Bloch state is perpendicular to the axis. In this sense, $H_{\rm Pre}$\ cannot be taken as $H_{0}$ with parameters $\left( \kappa
_{x},\kappa _{y},\kappa _{z}\right) =\left( 0,0,0\right) $, since $%
\left\vert \Phi (0)\right\rangle $\ is uncertain in practice due to the
degeneracy. However, we can consider the setup as $H_{\rm Pre}=H_{0}+\kappa _{x}(D^{\dag }+D)$ and $H_{\rm Pos}=H_{0}+\kappa
_{x}(D^{\dag }+D)+\rmi\kappa _{y}(D^{\dag }-D)$, and the initial state is given
as the ground state of the prequench Hamiltonian: $\left\vert \Phi
(0)\right\rangle =\left( \left\vert \psi _{n}^{+}\right\rangle -\left\vert
\psi _{n}^{-}\right\rangle \right) /\sqrt{2}$, under which we have%
\begin{eqnarray}
L\left( t\right) &=&|\left\langle \Phi (0)\right\vert \rme^{\rmi H_{\rm Pos}t}\left\vert \Phi (0)\right\rangle |^{2}  \nonumber \\
&=&\frac{2\kappa _{x}^{2}+\kappa _{y}^{2}+\kappa _{y}^{2}\cos \left( 2\sqrt{%
\kappa _{x}^{2}+\kappa _{y}^{2}}t\right) }{2\left( \kappa _{x}^{2}+\kappa
_{y}^{2}\right) }.  \label{LE_theo}
\end{eqnarray}%
We note that $L\left( t\right) $ oscillates with period $\tau =\pi /\sqrt{%
\kappa _{x}^{2}+\kappa _{y}^{2}}$\ and amplitude $\kappa _{y}^{2}/(\kappa
_{x}^{2}+\kappa _{y}^{2})$, which tends to the maximum $1$ in the limit of $%
\kappa _{x}^{2}\ll \kappa _{y}^{2}$. The same conclusion can be obtained
when we consider the case with replacing $\rmi\kappa _{y}(D^{\dag }-D)$\ by $%
\kappa _{z}p$\ in $H_{\rm Pos}$.

One of the purposes of this section is to analyze the mechanism of the proposed quench protocol. In the next section, we will discuss the possible simplification of the perturbation in equation (\ref{Hp}) and the realization of the string operator.

\section{Identification of the quantum phase}

\label{IV}

In this section, we analyze the perturbation term 
and try to propose a practical scheme to demonstrate the dynamic detection
of the phase diagram. We note that $p$ is a typical string
operator, which is a challenge to realize in experiment. We first consider
the possible realization of the perturbation in practice. The perturbation
term $H^{\prime }$ in equation (\ref{Hp}) commutes to the unperturbed Hamiltonian 
$H_{0}$, then the whole Hamiltonian $H_{0}+H^{\prime }$ is exactly solvable.
However, the operators in $H^{\prime }$ are $g$ dependent and need to be
deliberately designed in practice.

\subsection{Simplified perturbations}

\label{IVA}

\begin{figure}[t]
	\centering
	\includegraphics[width=0.48\textwidth]{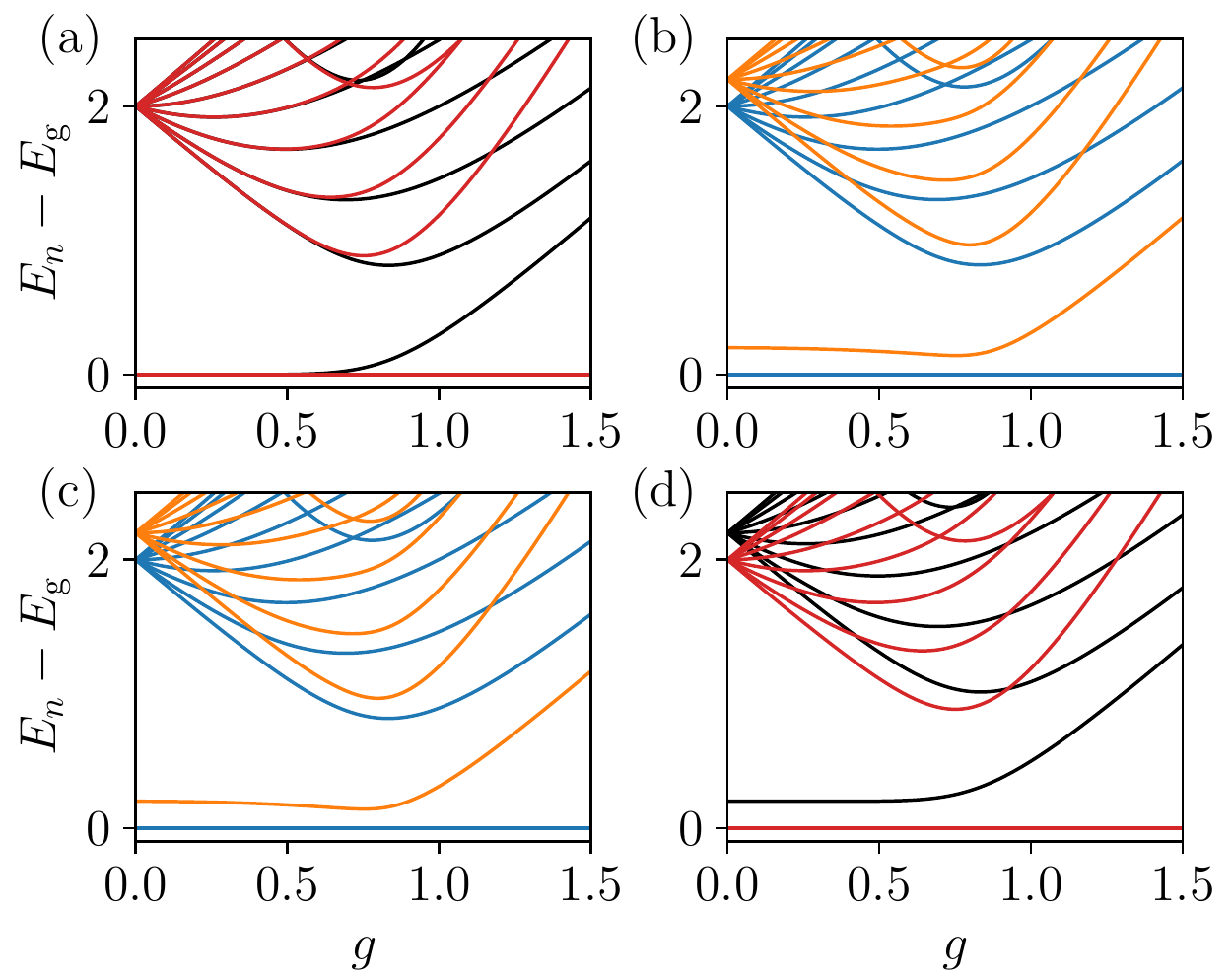}
	\caption{Spectrum of the low-lying eigenstates for Hamiltonian $H=H_{0}+H_{%
			\mathrm{S}}^{\prime }$ as a function of $g$ with parameters $\left( \protect%
		\kappa _{x},\protect\kappa _{y},\protect\kappa _{z}\right) $ (a) $\left(
		0,0,0\right) $, (b) $\left( 0.1,0,0\right) $, (c) $\left( 0,0.1,0\right) $
		and (d) $\left( 0,0,-0.1\right) $, obtained numerically through exact
		diagonalization. The red and black colors of the lines in (a) and (d) denote
		even and odd parities of the corresponding eigenstates, respectively, while
		the eigenstates of spectrum (b) and (c) are not the eigenstates of parity
		operator $p$. The spacing of the energy splits equal to $2\sqrt{\kappa_{x}^2+\kappa_{y}^2+\kappa_{z}^2}$ approximatively. 
		Here $E_{\mathrm{g}}$ is the ground-state energy. System
		parameters are $N=10$ and $J=1$.}
	\label{fig1}
\end{figure}

First, we consider a simplification of the perturbation in equation (\ref{Hp}),
that is, only the dominant terms of $H^{\prime }$ are taken into account: 
\begin{eqnarray}
H_{\mathrm{S}}^{\prime } &=&\kappa _{x}\sigma _{1}^{x}-\kappa
_{y}\prod\limits_{l=1}^{N-1}\left( -\sigma _{l}^{z}\right) \sigma
_{N}^{y}+\kappa _{z}\prod_{j=1}^{N}(-\sigma _{j}^{z})  \nonumber \\
&=&\kappa _{x}\sigma _{1}^{x}-\rmi\kappa _{y}p\sigma _{N}^{x}+\kappa _{z}p,
\label{H_s}
\end{eqnarray}%
which is equal to a small $g$ limit of $H^{\prime }$. We find that the local
operator $\sigma _{1 (N)}^{x}$ and the nonlocal operator $p$ are two
elemental actions of the perturbations. The advantages of considering this
perturbation are two folds: (i) $H_{\mathrm{S}}^{\prime }$ is independent of
the system parameter $g$, allowing us to implement the LE detection when the
system parameters are unknown. (ii) The form of $H_{\mathrm{S}}^{\prime }$ is
simpler and thus is more possible for experimental implementation. To see
the effects of the perturbation $H_{\mathrm{S}}^{\prime }$ on the spectrum
of $H_{0}$, in figure \ref{fig1}, we present the spectrum of the low-lying
eigenstates of the Hamiltonian $H=H_{0}+H_{\mathrm{S}}^{\prime }$ for
different parameters $\left( \kappa _{x},\kappa _{y},\kappa _{z}\right) $
and $g$. We can see that in the ordered phase, the perturbations with
different $\left( \kappa _{x},\kappa _{y},\kappa _{z}\right) $ all lead to
almost equal level splitting for a fixed $g$ and as $g$ varying, which
suggests that perturbation $H_{\mathrm{S}}^{\prime }$ has the same effect as
that in equation (\ref{Hp}). As a local perturbation, $\sigma _{1}^{x}$ can lift
the degeneracy [see figure \ref{fig1}(b)], which has the same effect as the
nonlocal case in figures \ref{fig1}(c) and \ref{fig1}(d). We also note that
perturbations $\kappa _{x}\sigma _{1}^{x}$ and $-\rmi\kappa _{y}p\sigma
_{N}^{x} $ both break the parity symmetry, while $\kappa _{z}p$ preserves it
[see the red (even parity) and black (odd parity) lines in figures \ref{fig1}%
(a) and \ref{fig1}(d)].

In contrast, we investigate the effect of another type of perturbation,
containing operator $\sigma_{N}^{y}$. Without loss of generality, we 
consider the case with Hamiltonian%
\begin{equation}
H=H_{0}+\sum_{j=1}^{N}\gamma _{j}\sigma _{j}^{y},
\end{equation}%
where $\gamma _{j}$\ is an arbitrary set of real numbers. Applying a set of
local transformation \cite{zhang2020ising} 
\begin{eqnarray}
\tau _{j}^{x} &=&\sigma _{j}^{x},  \nonumber \\
\tau _{j}^{y} &=&\eta _{j}^{+}\sigma _{j}^{y}-\eta _{j}^{-}\sigma _{j}^{z}, 
\nonumber \\
\tau _{j}^{z} &=&\eta _{j}^{+}\sigma _{j}^{z}+\eta _{j}^{-}\sigma _{j}^{y},
\end{eqnarray}%
with the factors $\eta _{j}^{+}=1/\sqrt{1+\gamma _{_{j}}^{2}}$ and $\eta
_{j}^{-}=\gamma _{_{j}}/\sqrt{1+\gamma _{_{j}}^{2}}$, we have%
\begin{equation}
H=-J\sum_{j=1}^{N-1}\tau _{j}^{x}\tau _{j+1}^{x}+\sum_{j=1}^{N}\sqrt{%
g^{2}+\gamma _{_{j}}^{2}}\tau _{j}^{z},
\end{equation}%
which is still a transverse field Ising chain since the new spin operators still satisfy the
Lie algebra commutation relations 
\begin{equation}
\left[ \tau _{j}^{\mu },\tau _{j}^{\nu }\right] =\sum_{\lambda
=x,y,z}2\rmi\epsilon ^{\mu \nu \lambda }\tau _{j}^{\lambda }.
\end{equation}%
Then weak perturbation $\kappa \sigma _{N}^{y}$ cannot lift the degeneracy
of $H_{0}$, since small derivation from uniform $g$ does not affect the
topological zero modes \cite{kitaev2001unpaired}.

\subsection{Realization of string operator action}

\label{IVB}

Second, we consider to realize the action of string operator $p$ by the time
evolution under a time-dependent local Hamiltonian%
\begin{equation}
H_{p}(t)=\mathfrak{g}(t)\sum_{l=1}^{N}\sigma _{l}^{z},
\end{equation}%
which describes the action of an extra time-dependent transverse field. Here
the coefficient is defined as%
\begin{equation}
\mathfrak{g}(t)=\left\{ 
\begin{array}{cc}
\frac{\pi }{2\Delta }, & 0<t\leq\Delta \\ 
0, & \rm otherwise%
\end{array}%
\right. .
\end{equation}%
After time $\Delta $, the effect of $H_{p}(t)$ on the degenerate state $%
\left\vert \psi _{n}\right\rangle =\alpha \left\vert \psi
_{n}^{+}\right\rangle +\beta \left\vert \psi _{n}^{-}\right\rangle $ can be
expressed as the time evolution operator%
\begin{eqnarray}
U(\Delta ) &=&\exp \left[ -\rmi\int_{0}^{\Delta }H_{p}(t)\mathrm{d}t\right] 
\nonumber \\
&=&\prod_{l=1}^{N}\exp \left( -\rmi\int_{0}^{\Delta }\frac{\pi }{2\Delta }%
\sigma _{l}^{z}\mathrm{d}t\right)  \nonumber \\
&=&\rmi^{N}\prod_{l=1}^{N}\left( -\sigma _{l}^{z}\right) .
\end{eqnarray}%
It indicates that the time evolution operator takes the role of the operator 
$p$, i.e.,%
\begin{equation}
U(\Delta )\left( \alpha \left\vert \psi _{n}^{+}\right\rangle +\beta
\left\vert \psi _{n}^{-}\right\rangle \right) =\rmi^{N}\left( \alpha
\left\vert \psi _{n}^{+}\right\rangle -\beta \left\vert \psi
_{n}^{-}\right\rangle \right) .
\end{equation}

\begin{figure*}[t]
	\centering
	\includegraphics[width=1\textwidth]{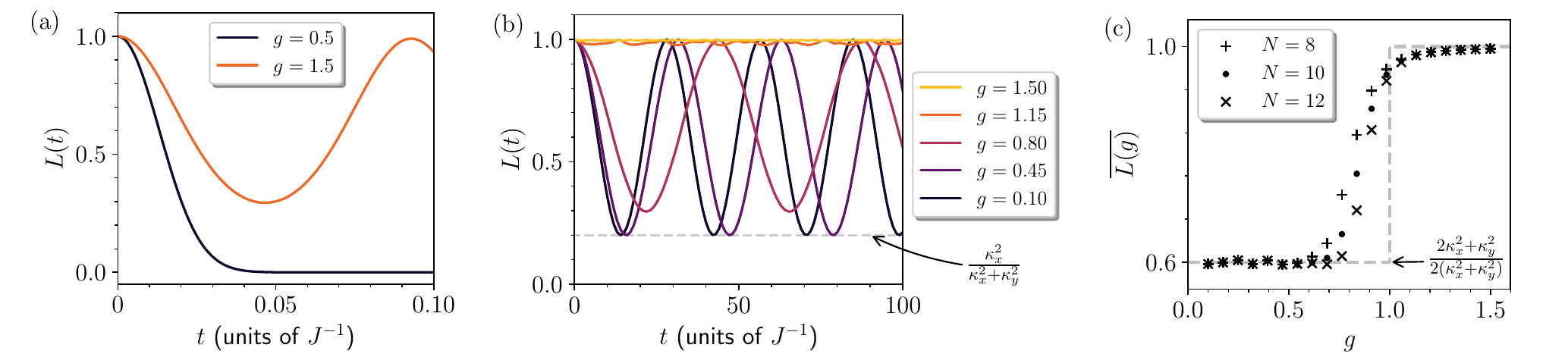}
	\caption{(a) Numerical results of LEs for the quench processes under the
		prequench and postquench Hamiltonians of equation (\protect\ref{HpreHposHp}), with
		parameters $g=0.5$ and $1.5$. Other parameters are $N=12$, $\protect\kappa%
		_{x}=0.05$ and $\Delta=0.10$. (b) Numerical results of LEs for the quench
		processes under Hamiltonians of equation (\protect\ref{HpreHposHS}) for different $g$
		values. The gray dashed line denotes the minimum value of LE obtained from
		equation (\protect\ref{LE_theo}). The system size is $N=12$. (c) Numerical
		results of average LEs as functions of $g$ [defined in equation (\protect\ref{avg_LE})] for $%
		N=8$, $10$ and $12$. The gray dashed line represents the ideal average LEs
		expected for large $N$ limits, and we set $T=500$.}
	\label{fig2}
\end{figure*}

To verify this result, we perform numerical simulation for a quench process
defined as%
\begin{eqnarray}
H_{\rm Pre} &=&H_{0}+\kappa _{x}\sigma _{1}^{x},  \nonumber \\
H_{\rm Pos} &=&H_{0}+H_{p}.  \label{HpreHposHp}
\end{eqnarray}%
The purpose of performing the prequench is to lift the degeneracy in the ordered phase, so that the eigenstates of the system become certain. The initial state is taken as the ground state of the prequench Hamiltonian. According to the previous results, it can be approximately expressed in the following form%
\begin{equation}
\left\vert \Phi (0)\right\rangle =\sin \frac{\theta }{2}\left\vert \psi _{%
\mathrm{g}}^{+}\right\rangle -\cos \frac{\theta }{2}\rme^{\rmi\varphi }\left\vert
\psi _{\mathrm{g}}^{-}\right\rangle .
\end{equation}%
Then in the ordered phase, the expected finial state is%
\begin{eqnarray}
\left\vert \Phi (\Delta )\right\rangle &=&U(\Delta )\left\vert \Phi
(0)\right\rangle  \nonumber \\
&=&\rmi^{N}\left( \sin \frac{\theta }{2}\left\vert \psi _{\mathrm{g}%
}^{+}\right\rangle +\cos \frac{\theta }{2}\rme^{\rmi\varphi }\left\vert \psi _{%
\mathrm{g}}^{-}\right\rangle \right) .
\end{eqnarray}%
It is expected that the LE obeys $L(\Delta )=\cos ^{2}\theta =\kappa
_{z}^{2}/\left\vert \mathbf{B}\right\vert ^{2}=0$. While in the disordered
phase, the ground state of $H_{0}$ is non-degenerate and is separated from
the excited state by an energy gap, then it is expected that $L(\Delta
)\approx1$. This scheme realizes the action of 
string operator $p$, i.e., the action of the first order term of time evolution operator $\rme^{-\rmi \kappa pt}$. The numerical results of LEs obtained by exact diagonalization
with system parameters $g=0.5$ and $1.5$\ are presented in figure \ref{fig2}%
(a), which are in accord with our analysis. Similarly, the actions of other string operators such as $\prod_{l<j} \left( -\sigma^{z}_{l} \right)  \sigma^{\alpha }_{j}\ (\alpha=x,y,z)$, can be realized by this scheme.

\subsection{Quantum phase transition}

\label{IVC}

According to the conclusions in sections \ref{III} and \ref{IVA}, it is
expected that the dynamics behavior of LE can be utilized to identify
different quantum phases when we consider a simplified version of the
perturbation in equation (\ref{H_s}). It should lead to the similar oscillatory 
behavior of LE described in equation (\ref{LE_theo}) in the ordered phase if one
implement the quench protocol. In contrast, such an oscillatory behavior of $%
L\left( t\right) $\ in the region of $g>1$ is absent, since the
non-degenerate eigenstates are not sensitive to the perturbation of $H_{%
\mathrm{S}}^{\prime }$. These features allow us to observe significantly
different dynamical behaviors in different quantum phases when the initial
state is chosen as any eigenstate of $H_{\rm Pre}$.

In the following, we consider the numerical simulation of quench process
under the Hamiltonian 
\begin{eqnarray}
H_{\rm Pre} &=&H_{0}+H_{\mathrm{S}}^{\prime }(\kappa_{x}=0.05,
\kappa_{y}=0,\kappa_{z}=0),  \nonumber \\
H_{\rm Pos} &=&H_{0}+H_{\mathrm{S}}^{\prime }(\kappa_{x}=0.05,
\kappa_{y}=0.1,\kappa_{z}=0),  \label{HpreHposHS}
\end{eqnarray}
where $H_{\mathrm{S}}^{\prime }(\kappa_{x}, \kappa_{y},\kappa_{z})$ is
defined in equation (\ref{H_s}). Here $\kappa_x$, $\kappa_y$ and $\kappa_z$ should be small compared to the energy scale of the system, so that $H_{\mathrm{S}}^{\prime}$ can be considered as a perturbation. We choose nonzero $\kappa_x$, $\kappa_y$ and zero $\kappa_z$ in order to compare with the analytical result in equation (\ref{LE_theo}). Similar results can be obtained for other $\mathbf{B}$ values, as long as the parameter vectors $\mathbf{B}=(\kappa_x, \kappa_y, \kappa_z)$ for the prequench and posquench Hamiltonians are not parallel or antiparallel to each other. The initial state is prepared as the ground
state of $H_{\rm Pre}$. In figure (\ref{fig2})(b), we presented
the LEs for different $g$, which are calculated by exact diagonalization. We
can see that the results are in accord with our predictions for both phases.
For small $g$, the minimum value of LE is the same as that of equation (\ref%
{LE_theo}), which is estimated under the complex version of quench term in
equation (\ref{Hp}).

This verifies that the proposed quench protocol can be utilized to identify
different quantum phase of the transverse field Ising chain. To compare with
the phase diagram in the thermodynamic limit, where the ordered phase and the disordered phase are separated by the critical point $g_{\mathrm{c}}=1$ \cite{pfeuty1970one}, we introduce the average LE in the time
interval $\left[ 0,T\right] $, which captures the change of the dynamics characteristic near the critical point, and can be used to infer the behavior of LE for a system with larger $N$. The average LE is defined as 
\begin{equation}
\overline{L(g)}=\frac{1}{T}\int_{0}^{T}L\left( t\right) dt,  \label{avg_LE}
\end{equation}
the value of which in the ordered phase can be estimated from equation (\ref%
{LE_theo}) in long-time limit, that is 
\begin{equation}
\overline{L(g<1)}=\frac{2\kappa _{x}^{2}+\kappa _{y}^{2}}{2\left( \kappa
_{x}^{2}+\kappa _{y}^{2}\right) }.
\end{equation}
While in the disordered phase, it is expected that $\overline{L(g>1)}=1$.
The numerical results of average LEs for different $g$ and $N$ are presented
in figure \ref{fig2}(c), which are obtained by exact diagonalization. We can
see that when the system size is larger, the average LE is closer to the
ideal values (gray dashed line) that are expected in the thermodynamic limit.

\subsection{Thermal state}

\label{IVD}

Now we discuss a possibility of applying the quench protocol to the thermal
state when $g<1$. In the previous section, we have known that for the Ising
chain with parameters $g<1$, the robust degeneracy occurs not only in the
ground states, but in all energy levels [see figure \ref%
{fig1}(a)]. In general, a thermal state of system $H_{0}$ with temperature $%
\beta ^{-1}$ can be written as $\rho _{0}=\rme^{-\beta H_{0}}/\Tr \rme^{-\beta H_{0}}$, which preserves no quantum information. However, the
robust degeneracy of the spectrum may enable a thermal state to preserve the
quantum information in each degenerate subspace, in the case that the
thermalization is induced by local perturbation from the environment.

Consider such a state as an initial state, with density matrix 
\begin{equation}
\rho =\frac{\sum_{n=1}^{2^{N-1}}\rme^{-\beta E_{n}}\left\vert \Phi
_{n}\right\rangle \left\langle \Phi _{n}\right\vert }{%
\sum_{n=1}^{2^{N-1}}\rme^{-\beta E_{n}}},
\label{rhi_mixed}
\end{equation}%
where%
\begin{equation}
\left\vert \Phi _{n}\right\rangle =\sin \frac{\theta _{n}}{2}\left\vert \psi
_{n}^{+}\right\rangle -\cos \frac{\theta _{n}}{2}\rme^{\rmi\varphi _{n}}\left\vert
\psi _{n}^{-}\right\rangle .
\end{equation}

\begin{figure}[t]
\centering
\includegraphics[width=1\textwidth]{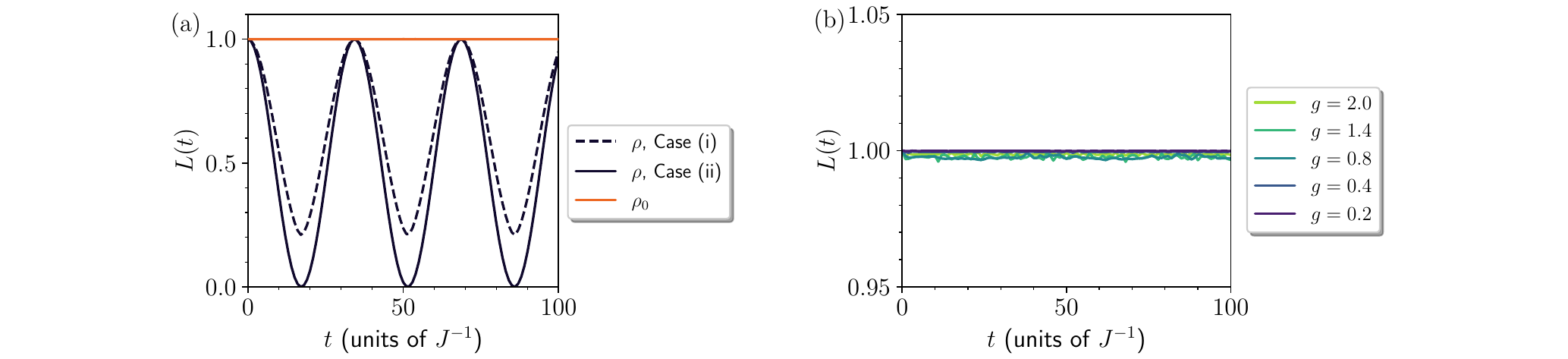}
\caption{(a) Numerical results of LEs for the quench processes under Hamiltonian in 
equation (\protect\ref{H_H0x}) and three different initial thermal states with $%
\protect\beta =1$. Here the legend ``$\rho$, Case (i)" denotes randomly distributed phase factors; ``$\rho$, Case (ii)" denotes fixed parity; and ``$\rho_{0}$" is the canonical ensemble distribution. 
The parameters of the system are $g=0.4$, $N=10$ and $%
\protect\kappa _{x}=0.1$. (b) Numerical results of LEs in different phases for the initial state of canonical ensemble distribution $\rho_{0}$. Other parameters are $\beta=1$, $N=10$ and $\kappa_{x}=0.1$.}
\label{fig3}
\end{figure}
At first, we estimate the dynamics of thermal state $\rho $ under the
quenched Hamiltonian 
\begin{equation}
H_{\mathrm{Pos}}=H_{0}+\kappa _{x}(D^{\dag }+D).  \label{H_H0D}
\end{equation}%
We choose this quenched Hamiltonian since the simplified version of the perturbation term is the simplest. Similar results can be obtained for the quench provided in equation (\ref{Hp}). 
In fact, we have 
\begin{equation}
\left\vert \left\langle \Phi _{n}\right\vert \rme^{-\rmi H_{\mathrm{Pos}%
}t}\left\vert \Phi _{n}\right\rangle \right\vert ^{2} 
=\cos ^{2}\left( \kappa _{x}t\right) +\sin ^{2}\left( \kappa _{x}t\right)
\sin ^{2}\theta _{n}\cos ^{2}\varphi _{n},
\label{overlap}
\end{equation}%
which is a periodic function of time. We consider two types of distribution of 
$\left\{ \theta _{n},\varphi _{n}\right\} $, which are encoded with
different informations. (i) Random distribution: $\left\{ \theta _{n}\right\} 
$ is taken as a random sample that is uniformly distributed over the
interval $[0,\pi )$ and $\left\{ \varphi _{n}\right\} =\left\{ 2\theta
_{n}\right\} $. Then the LE can be estimated by ignoring the Boltzmann
factor in high temperature and large $N$ limits, that is 
\begin{eqnarray}
L(t) &=&\frac{1}{2^{N-1}}\sum_{n=1}^{2^{N-1}}\left\vert \left\langle \Phi
_{n}\right\vert \rme^{-\rmi H_{\mathrm{Pos}}t}\left\vert \Phi _{n}\right\rangle
\right\vert ^{2}  \nonumber \\
&\approx &\frac{1}{\pi }\int_{0}^{\pi }d\theta \left[ \cos ^{2}\left( \kappa
_{x}t\right) +\sin ^{2}\left( \kappa _{x}t\right) \sin ^{2}\theta \cos
^{2}2\theta \right]   \nonumber \\
&=&\frac{1}{4}+\frac{3}{4}\cos ^{2}\left( \kappa _{x}t\right) .
\label{LE_case1}
\end{eqnarray}
(ii) Fixed parity: the thermal state consists of the levels with the same
parity, that is, $\theta _{n}=\pi $ and $\left\{ \varphi _{n}\right\} $ is
taken as a random sample in $[0,2\pi )$. Then we have 
\begin{eqnarray}
L(t) &\approx &\frac{1}{2\pi }\int_{0}^{2\pi }d\varphi \cos ^{2}\left(
\kappa _{x}t\right)   \nonumber \\
&=&\cos ^{2}\left( \kappa _{x}t\right) .  \label{LE_case2}
\end{eqnarray}%
For both cases, $L(t)$ are periodic functions but with different
amplitudes, and note that the minimal value for the latter is zero. 
When the posquench Hamiltonian is fixed, the difference of the amplitudes is originate from the initial states with parameters $\theta_{n}$ and $\varphi_{n}$, which may preserve different quantum informations, and reflect the coherence between the states with even and odd parity. When the initial state contains only one component of parity, the amplitude is $1$. 

In practice, based on the above analysis, one can consider the following
quenched Hamiltonian instead of equation (\ref{H_H0D}) 
\begin{equation}
H_{\mathrm{Pos}}=H_{0}+\kappa _{x}\sigma _{1}^{x},  \label{H_H0x}
\end{equation}%
and the definition of LE for density matrix is%
\begin{equation}
L(t)=\left[\Tr \sqrt{\sqrt{\rho (0)}\rho (t)\sqrt{\rho (0)}}\right]
^{2},  \label{LE_thermal}
\end{equation}%
which is also known as the Uhlmann fidelity \cite{uhlmann1976transition,
jacobson2011unitary}, characterizing the similarity between the initial
state $\rho (0)$ and evolved state $\rho (t)=\rme^{-\rmi H_{\mathrm{Pos}}t}\rho
(0)\rme^{\rmi H_{\mathrm{Pos}}t}$. The numerical results for the two types of
random initial states described above are presented in figure \ref{fig3}(a). As a
comparison, the numerical result for initial thermal state with canonical ensemble distribution $\rho
_{0}=\rme^{-\beta H_{0}}/\Tr \rme^{-\beta H_{0}}$ is also given. We can see
that for initial states $\rho $ of two cases, the LEs are close to the
results in equations (\ref{LE_case1}) and (\ref{LE_case2}), although the
definitions of LEs and the forms of the quench Hamiltonian are different.
For the initial thermal state $\rho _{0}$, the dynamics is not sensitive to
the perturbation in equation (\ref{H_H0x}). We can see that the amplitudes reflect  the coherence between the states with even and odd parity. The amplitude is zero for the canonical ensemble distribution $\rho_{0}$.  

\begin{figure}[t]
  \centering
	\includegraphics[width=1\textwidth]{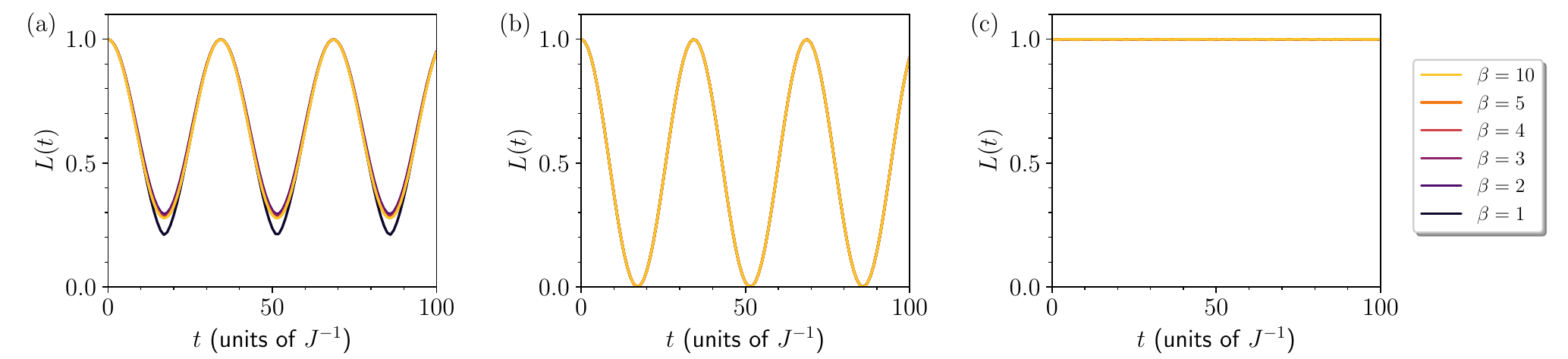}
  \caption{(a) Numerical results of LEs for different inverse temperatures $\beta$ and three different initial thermal states: (a) Case (i) randomly distributed phase factors. (b) Case (ii) with fixed parity. Here the same random distribution of $\left\{ \theta_{n} ,\  \varphi_{n} \right\}  $  are used for different $\beta$ to produce comparable results. (c) Canonical ensemble distribution $\rho_{0}$. The parameters of the system are $g=0.4$, $N=10$ and $\kappa_{x}=0.1$.}
    \label{figure_tem}
\end{figure}

Since the constructions of the thermal states for cases (i) and (ii) are based on the two-fold degenerate spectrum (see the definition in equation (\ref{rhi_mixed}), where $E_{n}=\varepsilon_{n}^{+}=\varepsilon_{n}^{-}$ is the energy for degenerate states $|\psi^{\pm}_{n} \rangle$), the definition of the thermal state in case (i) and (ii) is absent in the disordered phase, where the energy levels are not degenerate. Therefore, the quench dynamics in the disordered phase  can be discussed only for the initial state of canonical ensemble distribution $\rho_{0}$. To investigate the behavior of LE in the disordered phase, numerical calculations are carried out with the initial state $ \rho_{0}$ for different $g$, and the results are presented in figure \ref{fig3}(b). As we can see, the LEs are not sensitive to the variation of parameter $g$, meaning that the phase transition can not be detected for thermal ensembles by the quenched Hamiltonian considered in this paper. 

To see the temperature dependence, we carried out the numerical calculations. The results in figure \ref{figure_tem}(a) show that for case (i) with randomly distributed phase factors, the amplitudes of LEs are slightly different at different inverse temperature $\beta$. In figures \ref{figure_tem}(b) and \ref{figure_tem}(c), we can see that the LEs are both not sensitive to the change of temperature for case (ii) with fixed parity and canonical ensemble distribution $\rho_{0}$. 

The system size $N$ mainly affects the pseudo critical point $g=g_{\mathrm{pc}}$ ($g_{\mathrm{pc}}\rightarrow1$ when $N\rightarrow\infty$) for the transition between the degenerate and non-degenerate regions. Since the discussion on the thermal states focus on the degenerate region, the finite size effect can be avoided by choosing $g\ll 1$.

\section{Summary}

\label{V}

In summary, we have studied the consequence of the Hermitian nonlocal perturbation
term on the transverse field Ising chain. The Hermitian perturbations is more convenient for experimental implementation than the non-Hermitian method.  We proposed a pseudospin
description for the Hamiltonian with perturbation term. In this description,
the perturbation acts as an effective magnetic field, which lift the
degenerate spectrum of the Hamiltonian in the ordered phase. The identical split in each level enable the same oscillatory dynamics for the excited states as that for the ground state, which means that the quench protocol can be applied to the ground state, as well as the excited states. We have shown
that the string operator action can be realized via a quench process under
the local perturbations. As an application, it is demonstrated that any
ground states and excited states for the Hamiltonian with perturbation can
be selected for identifying the quantum phase, by adding the other
perturbation to trigger a quench and measuring the LE. Our method provides another option to identify the quantum phases, while it is failed for the thermal state. It is possible to be improved by seeking for other quench protocol, such as using other postquench Hamiltonian. It can  be applied to other model with open boundary condition, as long as the corresponding edge operator exists. Our work, including the numerical results of LEs for a small-size system, provides a possible realization of the nonlocal operation as well as alternative quench protocol
to detect the QPT. In addition, the result may shed light on the protocol of
quantum information processing based on nonlocal pseudospin as qubit.

\ack This work was supported by the National Natural Science
Foundation of China (under Grant No. 11874225).

\section*{Data availability statement}
All data that support the findings of this study are included within the article (and any supplementary files).

\providecommand{\newblock}{}

\end{document}